# Static conductivity of charged domain wall in uniaxial ferroelectric-semiconductors

E.A. Eliseev[1,2], A.N. Morozovska[1*], G.S. Svechnikov[1], Venkatraman Gopalan[3], V.Ya. Shur[4†]

[1] Institute of Semiconductor Physics, National Academy of Science of Ukraine,
41, pr. Nauki, 03028 Kiev, Ukraine

[2] Institute for Problems of Materials Science, National Academy of Science of Ukraine,
3, Krjijanovskogo, 03142 Kiev, Ukraine

[3] Department of Materials Science and Engineering, Pennsylvania State University,
University Park, Pennsylvania 16802, USA

[4] Institute of Physics and Applied Mathematics, Ural State University,
Ekaterinburg 620083, Russia

**Abstract**

Using Landau-Ginzburg-Devonshire theory we calculated numerically the static conductivity of both inclined and counter domain walls in the uniaxial ferroelectrics-semiconductors of *n*-type. We used the effective mass approximation for the electron and holes density of states, which is valid at arbitrary distance from the domain wall.

Due to the electrons accumulation, the static conductivity drastically increases at the inclined head-to-head wall by 1 order of magnitude for small incline angles $\theta \sim \pi/40$ by up 3 orders of magnitude for the counter domain wall ($\theta=\pi/2$). Two separate regions of the space charge accumulation exist across an inclined tail-to-tail wall: the thin region in the immediate vicinity of the wall with accumulated mobile holes and the much wider region with ionized donors. The conductivity across the tail-to-tail wall is at least an order of magnitude smaller than the one of the head-to-head wall due to the low mobility of holes, which are improper carries. The results are in qualitative agreement with recent experimental data for $LiNbO_3$ doped with MgO.

---

[*] Corresponding author, e-mail: morozo@i.com.ua
[†] Corresponding author, e-mail: vladimir.shur@usu.ru

# 1. Introduction

Conductive ferroelectric domain walls are very interesting for fundamental studies as well as promising for nanoelectronics development due to their nanosized width as well as the possibility to control their spatial location by external fields. In particular, Seidel et al [1] reported the observation of room-temperature electronic conductivity at ferroelectric domain walls in the insulating multiferroic $BiFeO_3$. The origin of the observed conductivity was probed using a combination of conductive atomic force microscopy, high-resolution transmission electron microscopy and first-principles density functional computations. Performed analyses revealed that the conductivity distribution correlates with structurally driven changes in both the electrostatic potential and the local electronic structure, which shows a decrease in the band gap at the domain wall.

Charged domain walls cannot be thermodynamically stable in ferroelectrics and ferroelectrics-semiconductors. However charged domain walls inevitably originate during the process of ferroelectric polarization reversal. During a real polarization reversal in a ferroelectric capacitor, the needle-like domains with charged domain walls arised at the polar surface move through the sample [2, 3, 4, 5]. The formation of the quasi-regular cogged charged domain wall and its expansion have been studied experimentally in $LiNbO_3$ under polarization reversal with uniform metal electrodes [4]. Domain wall pinning and bowing originate from defect centers [6]. Isolated wedge-shaped domains are formed under the charged SPM probe which then grow through the uniaxial ferroelectric of nano-, micro- or millimeter thickness acquiring an almost cylindrical shape or a slightly truncated cone [7, 8, 9, 10] or long needles [11, 12, 13, 14]. Note, that from one to three orders of magnitude increase of the bulk conductivity along the atrificially produced charged domain wall has been measured in single crystal of ferroelectric-semiconductor SbSJ [12].

Charged domain walls, shown in **Fig. 1a-d,** depending on the bound charge discontinuity at the wall (i.e. depending on the incline angle θ between the wall plane and polarization vector of the uniaxial ferroelectric), create strong electric fields, which in turn cause free charge accumulation across the wall and sharply increase the domain wall conductivity. When an inclined domain wall grows through the ferroelectric (as shown in **Fig. 1f**), it may become a conducting channel, and the strong increase of conductivity current will be registered by current scanning probe microscopy (CSPM), until the wall becomes uncharged again (as shown in **Fig. 1g**). Since the bound charge distribution is continuous across the uncharged 180° domain wall, such walls do not create any electric fields and naturally do not induce any redistribution of the free charges across the wall.

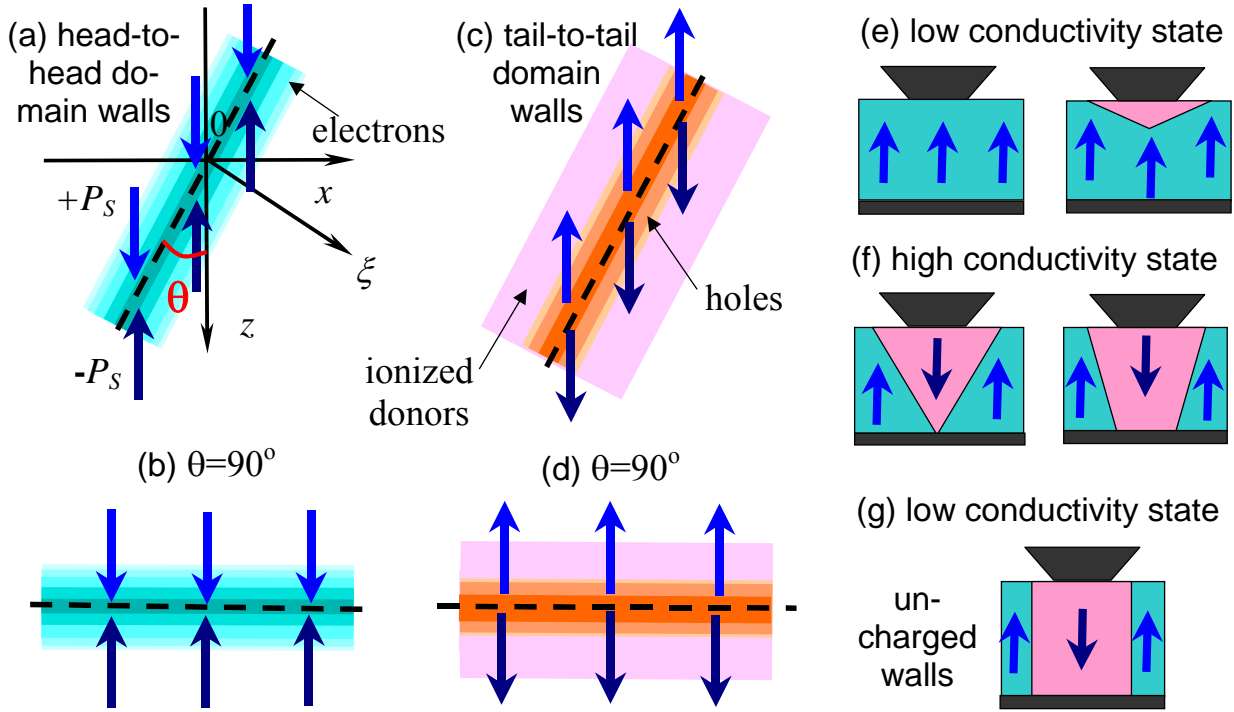

**Fig. 1.** (a-d) Sketch of the charged walls in the uniaxial ferroelectrics-semiconductors of *n*-type: inclined head-to-head (a), counter head-to-head (b), inclined tail-to-tail (c) and counter tail-to-tail domain walls, where the gradient colors indicate the free carrier concentration (electrons in the case (a,b) and donors+holes in the case (b,d)) increase at the domain wall vicinity. The incline angle of the domain wall is $\theta$. (e-g) When switching from state e to state g, the intermediate high conductivity state may appear due to the intergrowth of the charged domain walls during the local polarization reversal in the uniaxial ferroelectrics-semiconductors. External voltage is applied between the current scanning probe microscope tip (CSPM) and bottom electrode.

Analyses of the literature shows that the important problem of the charged domain wall conductivity was not enough studied theoretically. For instance, Guro et al.[15, 16] used Boltzmann approximation for dependence of holes and electrons on the electrostatic potential and consider only counter domain walls (for a detailed review see textbook of Fridkin [17]). However the studies consider the intrinsic semiconductor-ferroelectrics, while only the oversimplified estimations of the band bending (maximal potential value) and carrier concentrations near the surface with zero polarization are made for extrinsic semiconductor ferroelectrics with impurities.

Mokry et al.[18] consider an infinitely thin inclined domain wall without any internal structure of the screening and bound charge distribution. Concrete calculations are performed for the case when both bound charges and screening charges (proportional to the bound ones) are localized directly at the domain wall plane, while the self-consistent calculation of the screening charge distribution across the wall was not performed.

Using Landau theory, Gureev et al. [19, 20] considered the problem of the structure and energy of a charged 180° head-to-head domain wall. It was found that the scales controlling the wall structure can be very different from the Debye radius. Depending on the spontaneous polarization and the concentration of free carriers, these scales can be about the Thomas-Fermi screening length or about those typical for screening in nonlinear (Thomas-Fermi or Debye) regimes.

To summarize the brief overview, the conductivity distribution across a charged domain walls has not been calculated previously even for uniaxial ferroelectrics. This fact motivates our study: we calculated the static conductivity of both inclined and counter domain walls in the uniaxial ferroelectric-semiconductor using Landau-Ginzburg-Devonshire theory. We used the effective mass approximation for the electron and holes density of states, which is valid at arbitrary distance from the domain wall.

## 2. Problem statement

Let us consider head-to-head and tail-to-tail inclined wall in uniaxial ferroelectric-semiconductor doped with *n*-type impurity (e.g. LiNbO$_3$:Fe, Mg or LiTaO$_3$:Cr, etc). Sketch of the charged walls is shown in **Fig.1a-d.** The incline angle of the domain wall is regarded as $\theta$.

For the uniaxial ferroelectrics, the electric potential $\varphi(x, z)$ and ferroelectric polarization component $P_z(x, z)$ should be found from the Poisson equation:

$$\varepsilon_0 \left( \varepsilon_{33}^b \frac{\partial^2 \varphi}{\partial z^2} + \varepsilon_{11} \frac{\partial^2 \varphi}{\partial x^2} \right) = \frac{\partial P_z}{\partial z} - q\left( N_d^+(\varphi) + p(\varphi) - n(\varphi) - N_a^- \right), \tag{1a}$$

with boundary conditions of the potential vanishing far from the domain wall:

$$\varphi(r \to \infty) = 0, \quad \varphi(r \to -\infty) = 0. \tag{1b}$$

The charges are in the units of electron charge $q=1.6\times10^{-19}$ C, $\varepsilon_0=8.85\times10^{-12}$ F/m is the universal dielectric constant, $\varepsilon_{33}^b$ is the background dielectric permittivity of the ferroelectric, $\varepsilon_{33}^b$. Here ionized deep acceptors with field-independent concentration $N_a^-$ play the role of a background charge, ionized shallow donors and free holes and electrons equilibrium concentration are $N_d^+$, $p$ and $n$.

$$N_d^+(\varphi) = N_{d0}(1 - f(E_d - E_F - q\varphi)), \tag{2a}$$

$$p(\varphi) = \int_0^\infty d\varepsilon \cdot g_p(\varepsilon) f(\varepsilon - E_V + E_F + q\varphi)$$
$$\approx \left(\frac{m_p k_B T}{\hbar^2}\right)^{3/2} \frac{1}{\pi^2 \sqrt{2}} \frac{\sqrt{\pi}}{2} \left(-\text{Li}_{3/2}\left(-\exp\left(\frac{-q\varphi + E_V - E_F}{k_B T}\right)\right)\right), \tag{2b}$$

$$n(\varphi) = \int_0^\infty d\varepsilon \cdot g_n(\varepsilon) f(\varepsilon + E_C - E_F - q\varphi)$$
$$\approx \left(\frac{m_n k_B T}{\hbar^2}\right)^{3/2} \frac{1}{\pi^2 \sqrt{2}} \frac{\sqrt{\pi}}{2} \left(-\text{Li}_{3/2}\left(-\exp\left(\frac{q\varphi + E_F - E_C}{k_B T}\right)\right)\right). \tag{2c}$$

Where $N_{d0}$ is the donors concentration, $f(x) = \dfrac{1}{1 + \exp(x/k_B T)}$ is the Fermi-Dirac distribution function, $k_B = 1.3807 \times 10^{-23}$ J/K, $T$ is the absolute temperature. $E_F$ is the Fermi energy level, $E_d$ is the donor level, $E_C$ is the bottom of conductive band, $E_V$ is the top of the valence band (all energies are counted from the vacuum level). When the "bulk" density of states will be $g_n(\varepsilon) \approx \dfrac{\sqrt{2m_n^3 \varepsilon}}{2\pi^2 \hbar^3}$ and $g_p(\varepsilon) \approx \dfrac{\sqrt{2m_p^3 \varepsilon}}{2\pi^2 \hbar^3}$ in the effective mass approximation (usually $m_n \ll m_p$), one obtains the approximate equalities in Eqs.(2b,c), where $\text{Li}_n(z) = \sum_{k=1}^\infty \dfrac{z^k}{k^n}$ is the polylogarithmic function.

Due to the potential vanishing far from the wall (see Eq.(1b)) the condition should be valid:

$$N_a^- = N_{d0}^+ + p_0 - n_0, \tag{3}$$

where $N_{d0}^+ = N_{d0}(1 - f(E_d - E_F)) \equiv N_{d0} f(E_F - E_d)$, $p_0 = \int_0^\infty d\varepsilon \cdot g_p(\varepsilon) f(\varepsilon + E_F - F_V)$ and

$$n_0 = \int_0^\infty d\varepsilon \cdot g_n(\varepsilon) f(\varepsilon + E_C - F_F).$$

Polarization distribution satisfies LGD equation:

$$\alpha(T) P_z + \beta P_z^3 + \gamma P_z^5 - g\left(\frac{\partial^2 P_z}{\partial z^2} + \frac{\partial^2 P_z}{\partial x^2}\right) = -\frac{\partial \varphi}{\partial z} \tag{4a}$$

with the boundary conditions

$$P_z(r \to \infty) = P_S, \quad P_z(r \to -\infty) = -P_S \qquad (4b)$$

Domain wall plane is $z/x = -\cot\theta$ (see **Fig.1a**). Introducing new coordinate system, rotated around Y-axis on the angle $\theta$, and new variable:

$$\xi = x\cos\theta + z\sin\theta. \qquad (5)$$

Far from the crystal plate boundaries all the quantities depends only on $\xi$ and LGD Eq.(1a) and Poisson Eq.(4a) acquire the form of two coupled equations:

$$\alpha(T)P_z + \beta P_z^3 + \gamma P_z^5 - g\frac{\partial^2 P_z}{\partial \xi^2} = -\sin\theta\frac{\partial\varphi}{\partial\xi} \qquad (6a)$$

$$\varepsilon_0\left(\varepsilon_{33}^b \sin^2\theta + \varepsilon_{11}\cos^2\theta\right)\frac{\partial^2\varphi}{\partial\xi^2} = \sin\theta\frac{\partial P_z}{\partial\xi} - q\left(N_d^+(\varphi) + p(\varphi) - n(\varphi) - N_a^-\right), \qquad (6b)$$

With boundary conditions from Eq.(1b) and (4b) written as:

$$P_z(\xi \to \infty) = P_S, \quad P_z(\xi \to -\infty) = -P_S, \quad \varphi(\xi \to \infty) = 0, \quad \varphi(\xi \to -\infty) = 0 \qquad (6c)$$

Donor impact to the static conductivity can be neglected, since ions mobility (if any) are much smaller than the electron one. So, the static conductivity can be estimated as:

$$\sigma(\xi) = q(\eta_e n(\xi) + \eta_p p(\xi)). \qquad (7)$$

It is seen that it is coordinate dependent as proportional to the charge carrier concentration. Since usually $m_n \ll m_p$ (and therefore the mobility $\eta_e \gg \eta_p$) the most pronounced is the static electronic conductivity.

## 3. Results and discussion

Numerical solution of Eqs.(6) are shown in **Figs. 2-3** for the inclined head-to-head and tail-to-tail domain walls for **LiNbO₃** material parameters: $\varepsilon_{33}^b = 5$, $\varepsilon_{11} = 84$, $\varepsilon_{33} = 30$, $\alpha = -1.95\cdot 10^9$ m/F, $\beta = 3.61\cdot 10^9$ m$^5$/(C$^2$F), $\gamma = 0$; $g \sim 10^{-10}$ V·m$^3$/C. Spontaneous polarization $P_S = \sqrt{-\alpha/\beta} = 0.73$ C/m$^2$ and coercive field $E_{coers} = 2\sqrt{-\alpha^3/27\beta} = 5.5\ 10^8$ V/m, correlation length $r_c = \sqrt{-g/\alpha} \approx 0.4$ nm. Band gap is 4 eV, donors level is 0.1 eV deep, $m_n = 0.05 m_e$, $m_p = 5 m_e$, where $m_e$ is the mass of the free electron, and $N_{d0}^+ = 10^{23}$, $10^{24}$, $10^{25}$, $10^{26}$ m$^{-3}$ (without acceptors). Also we suppose that $\eta_e \approx 100\eta_p$, since $m_n \approx 0.01 m_p$ [21].

Dependencies of polarization $P_z(\xi)/P_S$, electric field $E_z(\xi)/E_{coer}$, potential $\varphi(\xi)$, concentrations of electrons, ionized donors and relative static conductivity $\sigma(\xi)/\sigma(\infty)$ on the distance $\xi/r_c$ from the wall plane was calculated for the inclined **head-to-head** domain wall with different slope angles $\theta = \pi/2, \pi/6, \pi/20, \pi/40, 0$ (see curves 1, 2, 3, 4, 5 in **Figs.2**). Holes concentration appeared less than $10^{-40} m^{-3}$ (i.e. they are absent near the wall). The uncharged wall is the thinnest; the charged counter wall with maximal bound charge is the thickest (**Fig.2a**). Correspondingly the electric field and potential created by the wall bound charges and screening carriers are the highest for the counter wall ($\theta = \pi/2$) with maximal bound $2P_S$; it decreases with the bound charge decrease, i.e. with $\theta$ decrease, since the bound charge is $2P_S \sin\theta$, and naturally vanishes at $\theta = 0$ (**Fig.2b,c**). The "net" electric field of the bound charge attracts free electrons (see accumulation region $|\xi| < 25 r_c$ in **Fig.2d**) and repulses ionized donors (see depletion region $|\xi| < 25 r_c$ in **Fig.2e**) from the charged wall region. The electron concentration is the highest for the counter wall ($\theta = \pi/2$); it decreases with the bound charge decrease (i.e. with $\theta$ decrease) vanishes at $\theta = 0$ (compare maximal values for different curves in **Fig.2d**). As a result of electron accumulation the static conductivity drastically increases at the wall: up 3 orders of magnitude for $\theta = \pi/2$ to 1 order for $\theta = \pi/40$ (**Fig.2e**). Donor impact to the static conductivity of the head-to-head domain walls can be neglected, since ions mobility (if any) are much smaller than the electron one.

Analyzing the results shown in **Fig.2c, d, e** we are lead to the following conclusions about the applicability of the most commonly used approximations for the charge carrier concentration across the **charged head-to-head** domain wall:

1) Debye approximation in Eqs.(2) that demands $|q\varphi| \ll k_B T$ becomes valid only very far ($|\xi| \gg 25 r_c$) from the charged domain wall in LiNbO$_3$, since $|q\varphi| < k_B T$ only at $|\xi| \gg 25 r_c$ even for $\theta = \pi/40$ (see **Fig.2c** and use $k_B T$ = 0.025 eV at room temperature).

2) Boltzmann approximation for electrons, $n(\varphi) \approx n_0 \exp(q\varphi/k_B T)$, is invalid in the immediate vicinity of charged domain walls ($|\xi| < 10 r_c$) allowing for their strong accumulation here. Approximation of a strongly degenerate electron gas, $n(\varphi) \approx \frac{(2m_n)^{3/2}}{3\pi^2 \hbar^3}(q\varphi + E_F - E_C)^{3/2}$, is valid in the immediate vicinity of the domain walls. Boltzmann approximation for holes, $p(\varphi) \approx p_0 \exp(-q\varphi/k_B T)$, is valid everywhere. Boltzmann approximation for donors, $N_d^+(\varphi) \approx N_{d0}^+ \exp(-q\varphi/k_B T)$, is valid in the vicinity of the domain wall ($|\xi| < 25 r_c$) (see **Fig.2c,d**).

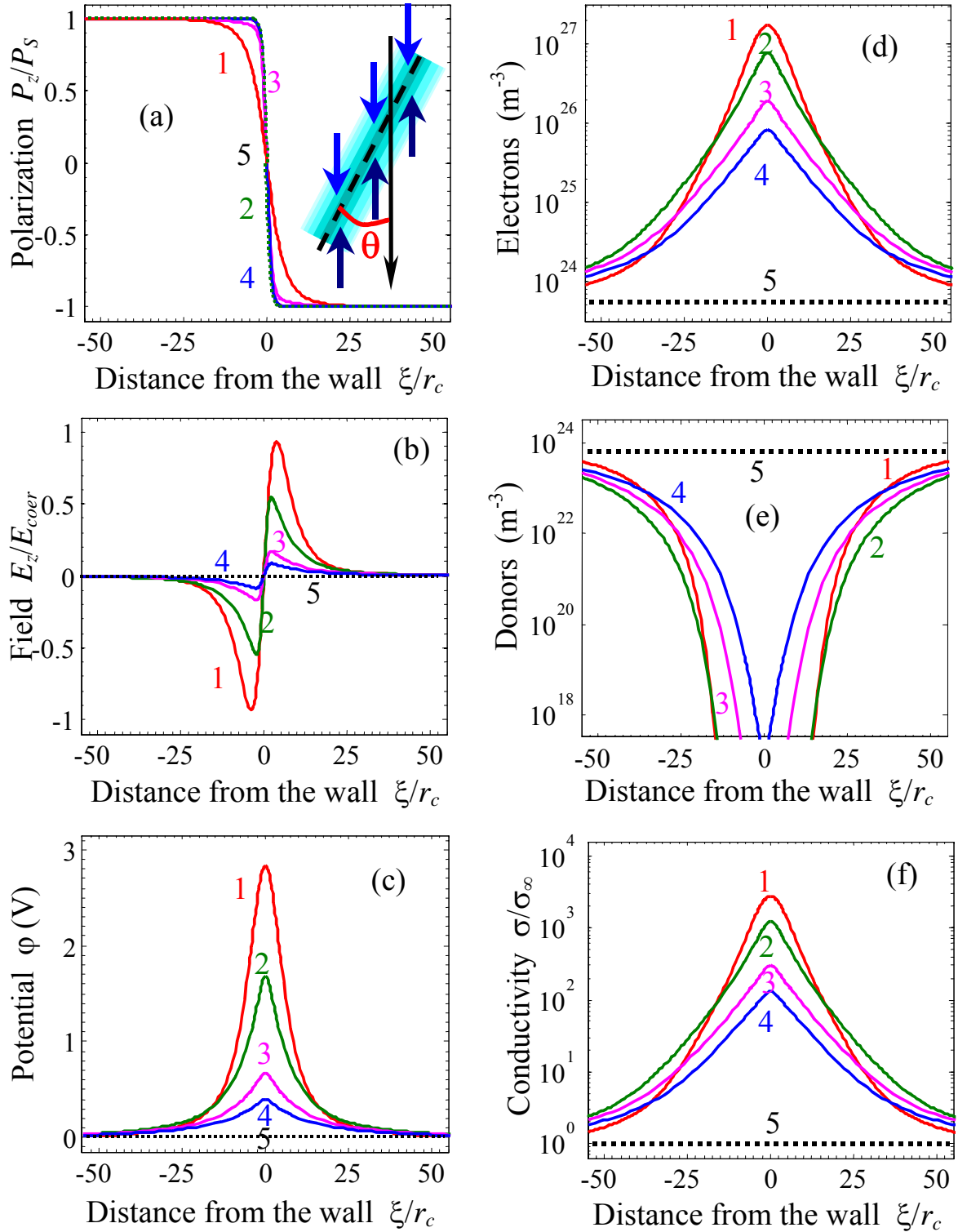

**Fig.2.** Dependencies of polarization $P_z(\xi)/P_S$ (a), field $E_z(\xi)/E_{coer}$ (b), potential $\varphi(\xi)$ (c), concentrations of electrons (d), ionized donors (e) and relative static conductivity $\sigma(\xi)/\sigma(\infty)$ (f) calculated for the inclined **"head-to-head"** domain wall with different slope angles

$\theta = \pi/2, \pi/6, \pi/20, \pi/40, 0$ (curves 1, 2, 3, 4, 5) and $N_{d0}^{+} = 10^{25}\,\text{m}^{-3}$. Holes concentration $<10^{-40}$ (i.e. they are absent near the wall). Material parameters correspond to LiNbO$_3$.

Dependencies of polarization, electric field, potential, concentrations of holes, electrons ionized donors and relative static conductivity on the distance $\xi/r_c$ from the wall plane was calculated for the inclined **tail-to-tail** domain wall with different slope angles $\theta = \pi/2, \pi/4, \pi/10, \pi/20, \pi/27, \pi/40, 0$ (see curves 1-7 in **Figs.3**). Note, that only the half of the tail-to-tail domain wall is shown in **Fig. 3** for the sake of clarity. Polarization of the uncharged wall saturates most quickly; the charged counter wall with maximal bound charge saturates most slowly, but the difference is small (compare different curves in **Fig.3a**). Electric field and potential created by the wall bound charges and screening carriers are the highest for the counter wall ($\theta = \pi/2$) with maximal bound $2P_S$; it decreases with the bound charge decrease, i.e. with $\theta$ decrease, since the bound charge is $2P_S \sin\theta$, and naturally vanishes at $\theta = 0$ (**Fig.3b,c**). The "net" electric field of the bound charge attracts holes in a very thin accumulation region $|\xi| < 5r_c$ (see solid curves in **Fig.3d,f**) and ionized donors (see thick depletion region $|\xi| < 100 r_c$ in **Fig.3e**) and repulses electrons from the charged wall region (see dashed curves in **Fig.3d,f**). The holes concentration is the highest for the counter wall ($\theta = \pi/2$); it decreases with the bound charge decrease (i.e. with $\theta$ decrease) vanishes at $\theta = 0$ (compare maximal values for different curves in **Fig.3d**). Electrons concentration appeared less than $10^{-40}\text{m}^{-3}$ near the wall, but dominates far from the wall as anticipated for *n*-type semiconductor (see dashed curves in **Figs.3d**). As a result of holes accumulation the static conductivity drastically increases at the wall: up 2 orders of magnitude (**Fig.3f**). Despite the qualitative similarity, the situation for the conductivity across the **tail-to-tail** wall is quantitatively different from the one for head-to-head wall: we see very thin accumulation region of mobile holes near the tail-to-tail wall and a very thick region of almost immobile donors that does not contribute to the wall conductivity, while the accumulation of mobile electrons is much thicker for the head-to-head wall (compare **Figs.2f** and **3f)**.

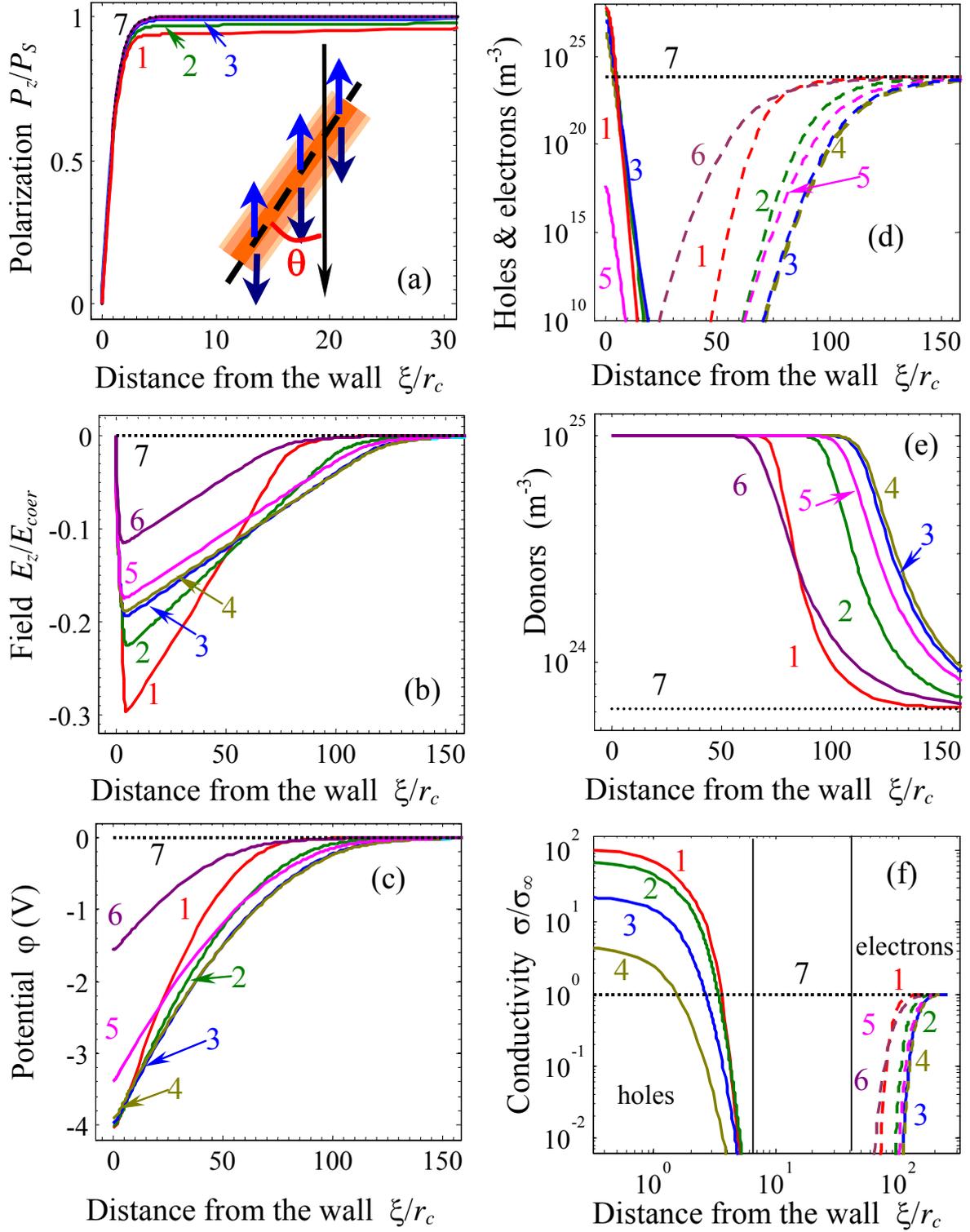

**Fig.3.** Dependencies of polarization $P_z(\xi)/P_S$ (a), field $E_z(\xi)/E_{coer}$ (b), potential $\varphi(\xi)$ (c), concentrations of holes (solid curves) and electrons (dashed curves) (d), ionized donors (e) and relative static conductivity $\sigma(\xi)/\sigma(\infty)$ (f) calculated for the inclined **"tail-to-tail"** domain wall with different slope angles $\theta = \pi/2, \pi/4, \pi/10, \pi/20, \pi/27, \pi/40, 0$ (curves 1, 2, 3, 4, 5, 6, 7)

and $N_{d0}^+ = 10^{25}\,\text{m}^{-3}$. Electron concentration $<10^{-40}$, i.e. they are absent near the wall. Material parameters correspond to LiNbO$_3$.

Analyzing the results shown in **Fig.3c, d, e** we lead to the conclusion about the applicability of the most commonly used approximations for the charge carrier concentration across the **charged tail-to-tail** domain wall:

1) Debye approximation in Eqs.(2) that demands $|q\varphi| \ll k_B T$ becomes valid only very far ($|\xi| \gg 25 r_c$) from the charged domain wall in LiNbO$_3$, since $|q\varphi| < k_B T$ only at $|\xi| \gg 25 r_c$ even for $\theta = \pi/40$ (see **Fig.3c**).

2) Boltzmann approximation for holes, $p(\varphi) \approx p_0 \exp(-q\varphi/k_B T)$, is invalid in the immediate vicinity of charged domain walls ($|\xi| < 10 r_c$) allowing for their strong accumulation here. Approximation of a strongly degenerate electron gas, $p(\varphi) \approx \dfrac{(2m_n)^{3/2}}{3\pi^2 \hbar^3}(-q\varphi - E_F + E_V)^{3/2}$, is valid in the immediate vicinity of the domain walls. Boltzmann approximation for electrons, $n(\varphi) \approx n_0 \exp(+q\varphi/k_B T)$, is valid near the wall. Full ionization of donors, $N_d^+(\varphi) \approx N_{d0}^+$, is valid in the vicinity of the domain wall ($|\xi| < 25 r_c$) (see **Fig.3c,d**).

Dependencies of polarization, electric field, potential, concentrations of electrons, holes, ionized donors and relative static conductivity vs. the distance $\xi/r_c$ from the wall plane was calculated for the limiting case of the counter domain walls (see **Fig. 4,5**).

It can be seen from **Fig. 4a-c** calculated for the head-to-head wall, that profiles of polarization, potential and electric field across the wall are practically independent on $N_{d0}^+$, since the screening is dominated by electrons (**Fig.4d**) and donor level is filled (concentration of ionized donors is small, see **Fig. 4e**) near the "head-to-head" wall (holes are absent everywhere). As a result of electron accumulation the static conductivity drastically increases at the wall (see **Fig.4f**).

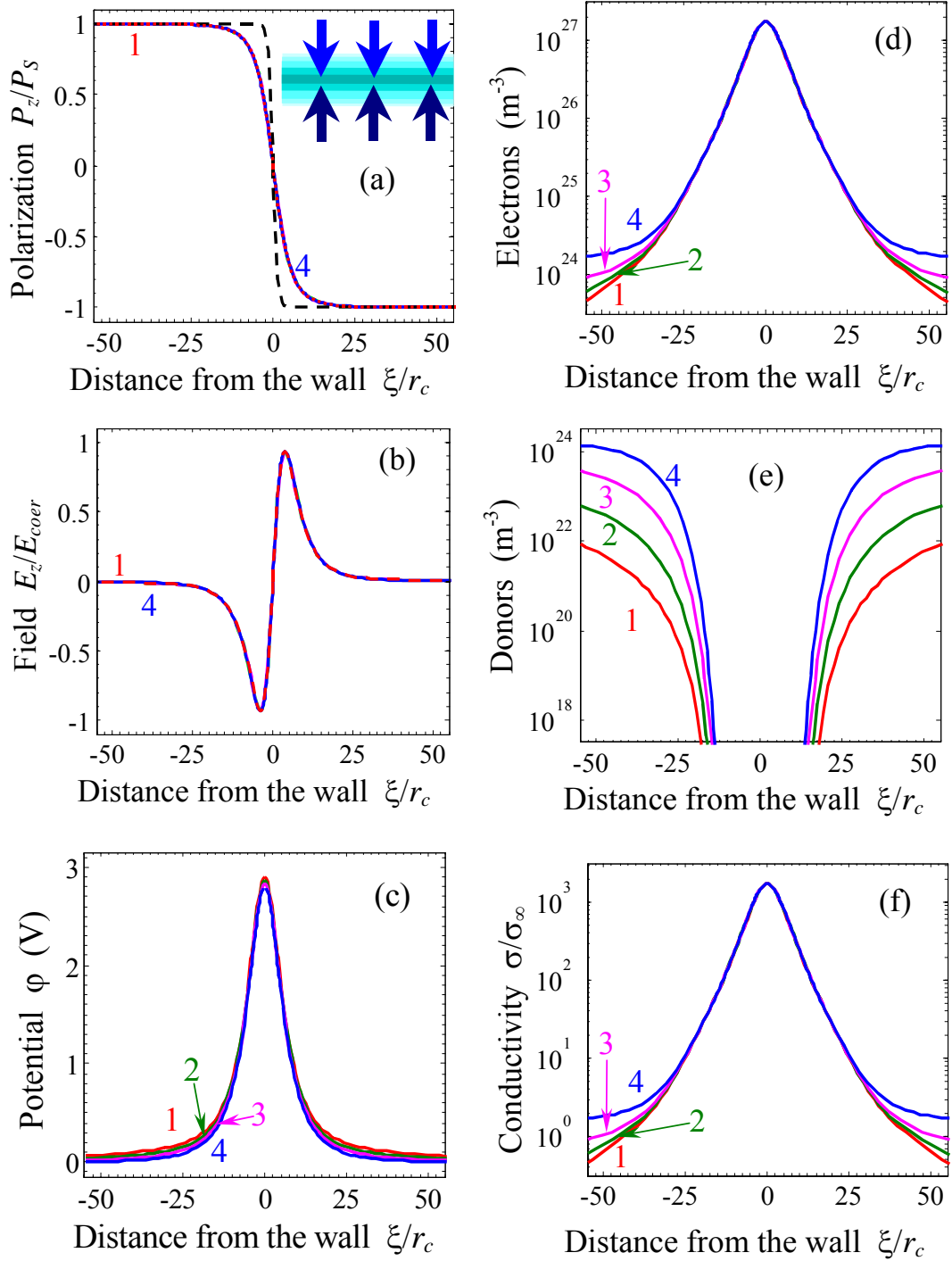

**Fig.4.** Dependencies of polarization $P_z(\xi)/P_S$ (a), field $E_z(\xi)/E_{coer}$ (b), potential $\varphi(\xi)$ (c), and concentrations of (d) electrons, (e) ionized donors and (f) relative static conductivity $\sigma(\xi)/\sigma(\infty)$ calculated for the counter **"head-to-head"** domain wall ($\theta = \pi/2$) and different $N_{d0}^{+} = 10^{23}$, $10^{24}$, $10^{25}$, $10^{26}$ m$^{-3}$ (curves 1, 2, 3, 4). Dashed curve in (a) is the profile of 180-degree uncharged domain wall. Holes concentration <$10^{-40}$ (i.e. they are absent near the wall). Material parameters correspond to LiNbO$_3$.

In contrast to the head-to-head walls, the profiles of polarization, potential and electric field across the counter **tail-to-tail** domain walls essentially depends on donor concentration $N_{d0}^+$, since here the negative bound charges are accumulated at the wall, which have to be screened by holes and ionized donors (see **Fig.5a-c**). Note, that only the half of the tail-to-tail domain wall is shown in **Fig. 5** for the sake of clarity. Since the equilibrium concentration of holes (improper carriers) is very small for the considered donor-type ferroelectric in comparison with the electrons, it should be enhanced near the tail-to-tail wall by either direct transition of electrons through the band gap or by donor ionization. However, the holes concentration increase is very limited by the donor ionization, as a result the structure of the counter tail-to-tail wall is completely different from the structure of the head-to-head one (compare **Fig.5a** with **Fig.4a**). Actually, one can see two separate regions of the space charge accumulation: the thin region in the immediate vicinity of the counter wall with accumulated holes and the much wider region with ionized donors (**Fig.5d,e,f**).

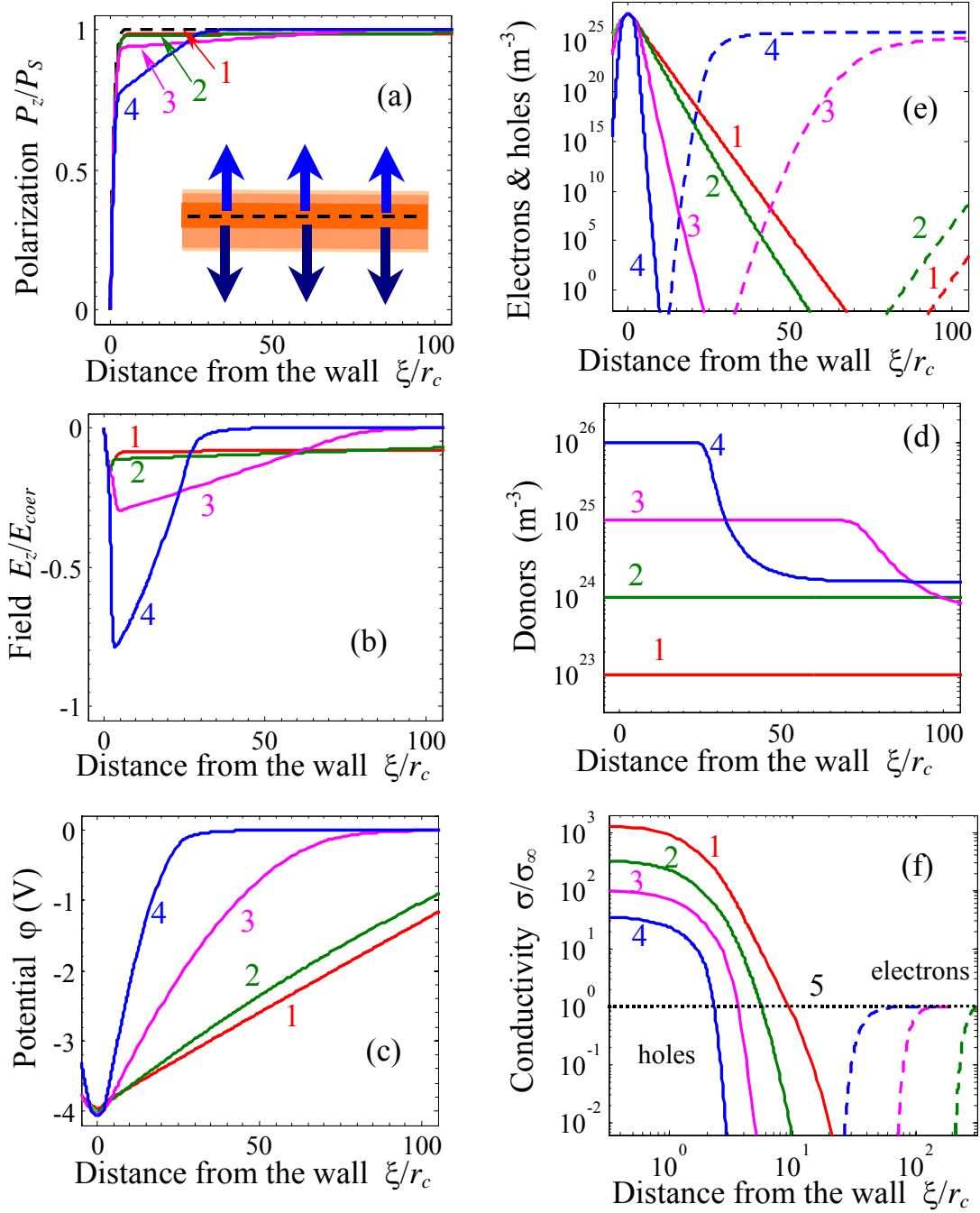

**Fig.5.** Dependencies of polarization $P_z(\xi)/P_S$ (a), field $E_z(\xi)/E_{coer}$ (b), potential $\varphi(\xi)$ (c), concentrations of (d) electrons (dashed curves) and holes (solid curves), (e) ionized donors and (f) relative static conductivity $\sigma(\xi)/\sigma(\infty)$ for a counter **"tail-to-tail"** domain wall ($\theta = \pi/2$) calculated for different $N_{d0}^+ = 10^{23}, 10^{24}, 10^{25}, 10^{26}$ m$^{-3}$ (curves 1, 2, 3, 4). Dashed curve in (a) is the profile of 180-degree domain wall (uncharged). Material parameters correspond to LiNbO$_3$.

Dependence of the static conductivity at the domain wall plane $\xi=0$, halfwidth at half maximum (FWHM) of polarization profile and mobile screening charges (electrons and holes) on the wall incline angle $\theta$ are compared in **Fig.6** for head-to-head and tail-to tail domain walls.

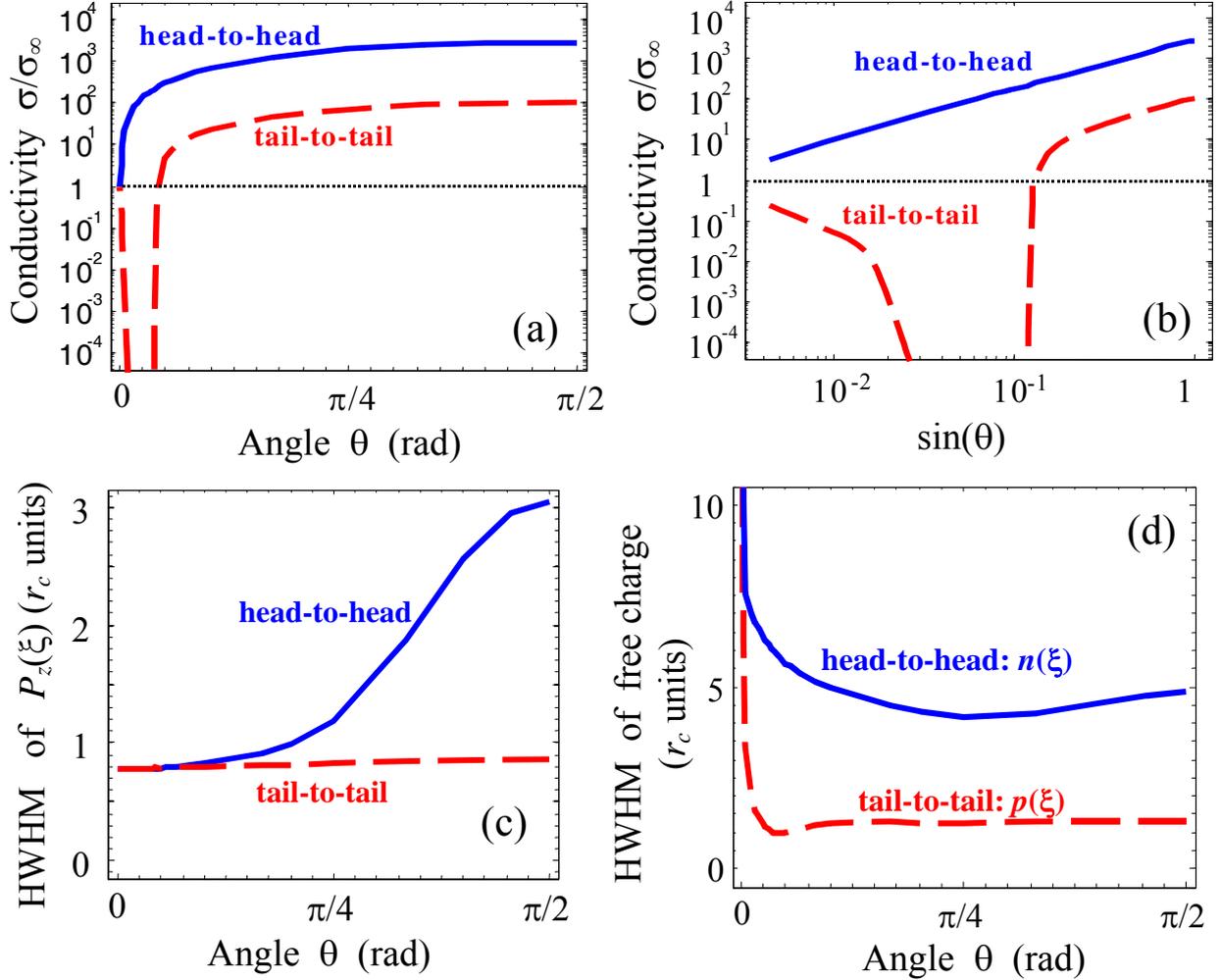

**Fig.6.** Dependence of the static conductivity at the domain wall plane $\xi=0$ (a – log-linear scale, b – log-log scale), halfwidth at half maximum (HWHM) of polarization profile (c) and mobile screening charges (d) on the wall incline angle $\theta$ for head-to-head (solid curves) and tail-to tail domain walls (dashed curves). Immobile ionized donors $\theta$–dependence is shown by dotted curve.

It can be seen from the **Figs.6a,b** that the static conductivity of the head-to-head wall is much higher (~30 times) than the one of the tail-to-tail wall for the considered $n$-type semiconductor-ferroelectric. Actually, the bound charge $+2P_S\sin\theta$ of the head-to-head wall is screened by the majority carriers – electrons, whose mobility and average concentration are much higher than for the minority carriers – "heavy" holes, which screen the bound charge $-2P_S\sin\theta$ of the tail-to-tail wall. It can be seen from the **Fig.6b** (plotted in double logarithmic scale) that the conductivity of head-to head wall is linearly proportional to $\sin\theta$. The conductivity of tail-to tail wall is linearly proportional to $\sin\theta$ except the region of small $\sin\theta < 0.1$, where the dip unexpectedly appears. The dip originated from the fact that amount of holes drastically decreases in the region $0.01 < \sin\theta < 0.1$, and almost immobile ionized acceptors performed the screening of the bound charge $-2P_S \sin\theta$.

It can be seen from the **Fig.6c** that the halfwidth of polarization profile of the head-to-head wall is only slightly higher (from 1 – 1,5 times $\theta < \pi/4$ at up to 3 times at $\theta \to \pi/2$) than the one of the tail-to-tail wall. The halfwidth of the screening electrons distribution across the head-to-head wall is always several times higher than the halfwidth of the screening holes distribution across the tail-to-tail wall, except the limit of the uncharged wall $\theta \to 0$ (see **Fig.6d**). The halfwidth of the screening charge depends on the wall incline angle $\theta$ non-monotonically. At very small angles ($\sin\theta << 0.1$) the wall bound charge becomes rather small and the screening carriers accumulation diffuses and becomes faint, as the result the halfwidth drastically increases. With $\theta$ increase the plateau (for the tail-to-tail wall) or very broad minimum (for the head-to-head wall) appears at $\theta \sim \pi/4$. With further $\theta$ increase from $\pi/4$ to $\pi/2$ the halfwidth of the head-to-head wall slightly increases.

**To summarize**, the structure of the screening charges distribution and static conductivity across the charged *inclined* head-to-head and tail-to-tail domain walls are very different in the *n*-type semiconductor-ferroelectrics.

1) Mobile electrons are accumulated in the vicinity of the head-to-head wall, which screen its bound charge $+2P_S \sin\theta$. The electric field and potential created by the wall bound charges and screening electrons (proper carriers) are the highest for the counter wall (incline angle $\theta = \pi/2$) with maximal bound $2P_S$; it decreases with the bound charge decrease, i.e. with $\theta$ decrease, since the bound charge is $2P_S \sin\theta$, and naturally vanishes at $\theta = 0$. As a result of electron accumulation the static conductivity drastically increases at the wall: up 3 orders of magnitude for $\theta = \pi/2$ to 1 order for $\theta = \pi/40$.

2) Two separate regions of the space charge accumulation exist across a tail-to-tail wall: the thin region in the immediate vicinity of the wall with accumulated mobile holes and the much wider region with ionized donors. Donor impact to the static conductivity of the domain walls can be neglected, since ions mobility (if any) are much smaller than the electron one. The conductivity across the tail-to-tail wall is at least an order of magnitude smaller than the one of the head-to-head wall due to the low mobility of holes, which are the improper carries.

3) Numerical results are compared with the model Boltzmann approximation for electrons and degenerated gas one. We have shown that Boltzmann approximation is valid far from the charged wall plane, while the degenerated gas one is valid specifically at the wall plane.

4) The high conductivity state may appear due to the intergrowth of the charged domain walls during the local polarization reversal in uniaxial ferroelectrics-semiconductors (**Fig.1d**). The result is in qualitative agreement with recent experimental data for $LiNbO_3$ doped with MgO [22].

**Acknowledgements**

EEA, ANM and GSV acknowledges financial of NAS Ukraine and Ukraine State Committee on Science, Innovation and Information (UU30/004). Research was sponsored in part by (VG, GSV, EEA and ANM) by National Science Foundation (DMR-0908718 and DMR-0820404). EEA, ANM and GSV acknowledge user agreement with CNMS N UR-08-869.

**Appendix A.**

**BA approximation.** In the Boltzmann approximation (**BA**), the Fermi-Dirac distribution function can be approximated as $f(x) \approx \exp(-x/k_B T)$, and the concentrations (2) acquire the form:

$$N_d^+(\varphi) \approx N_{d0}^+ \exp\left(\frac{-q\varphi}{k_B T}\right), \qquad p(\varphi) \approx p_0 \exp\left(\frac{-q\varphi}{k_B T}\right), \qquad n(\varphi) \approx n_0 \exp\left(\frac{q\varphi}{k_B T}\right). \qquad (A.1)$$

**DEG approximation.** For a strongly degenerate electron gas (**DEG**), the Fermi-Dirac distribution function can be approximated by a step function, $f(x) \approx \theta(x/k_B T)$, and the concentrations (2) acquire the form:

$$N_d^+(\varphi) = \frac{N_{d0}}{1 + \exp\left(-\dfrac{E_d - E_F - q\varphi}{k_B T}\right)}, \qquad (A.2a)$$

$$p(\varphi) \approx \frac{(2m_p)^{3/2}}{3\pi^2 \hbar^3}(-q\varphi + E_V - E_F)^{3/2}, \qquad n(\varphi) \approx \frac{(2m_n)^{3/2}}{3\pi^2 \hbar^3}(q\varphi + E_F - E_C)^{3/2}. \qquad (A.2b)$$